\documentclass[12pt]{article}
\usepackage{latexsym,amssymb,amsfonts,amsmath}
\usepackage[dvips]{graphicx}
\newlength{\extraspace}
\newlength{\extraspaces}
\newcommand{\be}{\begin{equation}
\addtolength{\abovedisplayskip}{\extraspaces}
\addtolength{\belowdisplayskip}{\extraspaces}
\addtolength{\abovedisplayshortskip}{\extraspace}
\addtolength{\belowdisplayshortskip}{\extraspace}}
\newcommand{\ee}{\end{equation}}

\newcommand{\bea}{\begin{eqnarray}
\addtolength{\abovedisplayskip}{\extraspaces}
\addtolength{\belowdisplayskip}{\extraspaces}
\addtolength{\abovedisplayshortskip}{\extraspace}
\addtolength{\belowdisplayshortskip}{\extraspace}}
\newcommand{\eea}{\end{eqnarray}}

\newcommand{\newsection}[1]{
\vspace{15mm}
\pagebreak[3]
\addtocounter{section}{1}
\setcounter{equation}{0}
\setcounter{subsection}{0}
\setcounter{footnote}{0}
\begin{flushleft}
{\Large\bf \thesection. #1}
\end{flushleft}
\nopagebreak
\medskip
\nopagebreak}

\def\Tr{{\rm Tr}}
\def\re{{\rm Re}}
\def\im{{\rm Im}}

\begin{document}

\addtolength{\baselineskip}{.8mm}

{\thispagestyle{empty}


\begin{center}
\vspace*{1.0cm}
{\large\bf CAN THE POMERON BE DERIVED FROM A EUCLIDEAN--MINKOWSKIAN DUALITY?}
\\
\vspace*{1.0cm}
{\large Enrico Meggiolaro\footnote{E--mail:
enrico.meggiolaro@df.unipi.it} }\\
\vspace*{0.5cm}{\normalsize
{Dipartimento di Fisica, Universit\`a di Pisa,
and INFN, Sezione di Pisa,\\
Largo Pontecorvo 3,
I--56127 Pisa, Italy.}}\\
\vspace*{2cm}{\large \bf Abstract}
\end{center}

\noindent
After a brief review, in the first part, of some relevant analyticity and
crossing--symmetry properties of the correlation functions of two Wilson loops
in QCD, when going from Euclidean to Minkowskian theory, in the second part we
shall see how these properties can be related to the still unsolved problem of
the asymptotic $s$--dependence of the hadron--hadron total cross sections.
In particular, we critically discuss the question if (and how) a
{\it pomeron}--like behaviour can be derived from this
Euclidean--Minkowskian duality.
\\
\\
}
\newpage

\newsection{Loop--loop scattering amplitudes}

\noindent
Differently from the parton--parton scattering amplitudes, which are known to
be affected by infrared (IR) divergences, the elastic scattering amplitude of
two colourless states in gauge theories, e.g., two $q \bar{q}$ meson states,
is expected to be an IR--finite physical quantity \cite{BL}.
It was shown in Refs. \cite{DFK,Nachtmann97,BN} (for a review see Refs.
\cite{Dosch,pomeron-book}) that the high--energy meson--meson elastic
scattering amplitude can be approximately reconstructed in two steps:
i) one first evaluates, in the functional--integral approach, the high--energy
elastic scattering amplitude of two $q \bar{q}$ pairs (usually called
{\it dipoles}), of given transverse sizes $\vec{R}_{1\perp}$ and
$\vec{R}_{2\perp}$ and given longitudinal--momentum fractions $f_1$ and $f_2$
of the two quarks in the two dipoles respectively;
ii) one then averages this amplitude over all possible values of
$\vec{R}_{1\perp}, f_1$ and $\vec{R}_{2\perp}, f_2$ with two proper squared
wave functions $|\psi_1 (\vec{R}_{1\perp},f_1)|^2$ and
$|\psi_2 (\vec{R}_{2\perp},f_2)|^2$, describing the two interacting mesons.
\footnote{One can also
take, for simplicity, the longitudinal--momentum fractions $f_1$ and $f_2$ of
the two quarks in the two dipoles (and, therefore, also the
longitudinal--momentum fractions $1-f_1$ and $1-f_2$ of the two antiquarks
in the two dipoles) to be fixed to $1/2$: this is known to be a good
approximation for hadron--hadron interactions (see Refs.
\cite{Dosch,pomeron-book} and references therein).}
(For the treatment of baryons, a similar, but, of course, more involved,
picture can be adopted, using a genuine three--body configuration or,
alternatively and even more simply, a quark--diquark configuration: we refer
the interested reader to the above--mentioned original references
\cite{DFK,Nachtmann97,BN,Dosch,pomeron-book}.)

The high--energy elastic scattering amplitude of two dipoles is
governed by the (properly normalized) correlation function of two Wilson loops
${\cal W}_1$ and ${\cal W}_2$, which follow the classical straight lines for
quark (antiquark) trajectories:
\be
{\cal M}_{(ll)} (s,t;\vec{R}_{1\perp},f_1,\vec{R}_{2\perp},f_2) \equiv
-i~2s \displaystyle\int d^2 \vec{z}_\perp
e^{i \vec{q}_\perp \cdot \vec{z}_\perp}
\left[ {\langle {\cal W}_1 {\cal W}_2 \rangle \over
\langle {\cal W}_1 \rangle \langle {\cal W}_2 \rangle} -1 \right] ,
\label{scatt-loop}
\ee
where $s$ and $t = -|\vec{q}_\perp|^2$ ($\vec{q}_\perp$ being the transferred
momentum) are the usual Mandelstam variables.
The expectation values $\langle {\cal W}_1 {\cal W}_2 \rangle$,
$\langle {\cal W}_1 \rangle$, $\langle {\cal W}_2 \rangle$ are averages
in the sense of the QCD functional integrals
and the Wilson loops ${\cal W}_1$ and ${\cal W}_2$ are so defined:
\be
{\cal W}^{(T)}_{1,2} \equiv
{1 \over N_c} \Tr \left\{ {\cal P} \exp
\left[ -ig \displaystyle\oint_{{\cal C}_{1,2}} A_\mu(x) dx^\mu \right]
\right\} ,
\label{QCDloops}
\ee
where ${\cal P}$ denotes the {\it path ordering} along the given path
${\cal C}$ and $A_\mu = A_\mu^a T^a$;
${\cal C}_1$ and ${\cal C}_2$ are two rectangular paths which
follow the classical straight lines for quark [$X_{q}(\tau)$, forward in
proper time $\tau$] and antiquark [$X_{\bar{q}}(\tau)$, backward in $\tau$]
trajectories, i.e.,
\bea
{\cal C}_1 &:&
X_{1q}^\mu(\tau) = z^\mu + {p_1^\mu \over m} \tau + (1-f_1) R_1^\mu ~~,~~
X_{1\bar{q}}^\mu(\tau) = z^\mu + {p_1^\mu \over m} \tau - f_1 R_1^\mu ,
\nonumber \\
{\cal C}_2 &:&
X_{2q}^\mu(\tau) = {p_2^\mu \over m} \tau + (1-f_2) R_2^\mu ~~,~~
X_{2\bar{q}}^\mu(\tau) = {p_2^\mu \over m} \tau - f_2 R_2^\mu ,
\label{traj}
\eea
and are closed by straight--line paths at proper times $\tau = \pm T$, where
$T$ plays the role of an IR cutoff \cite{Verlinde,Meggiolaro02}, which can
and must be removed in the end ($T \to \infty$).
Here $p_1$ and $p_2$ are the four--momenta of the two dipoles, taken for
simplicity with the same mass $m$, moving (in the center--of--mass system)
with speed $V$ and $-V$ along, for example, the $x^1$--direction:
\be
p_1 =
m \left( \cosh {\chi \over 2},\sinh {\chi \over 2},\vec{0}_\perp \right) ,~~~
p_2 =
m \left( \cosh {\chi \over 2},-\sinh {\chi \over 2},\vec{0}_\perp \right) .
\label{p1p2}
\ee
Here $\chi = 2~{\rm arctanh} V$ is the hyperbolic angle between the
two trajectories $1q$ and $2q$, i.e., $p_1 \cdot p_2 = m^2 \cosh\chi$.
Moreover, $R_1 = (0,0,\vec{R}_{1\perp})$, $R_2 = (0,0,\vec{R}_{2\perp})$
and $z = (0,0,\vec{z}_\perp)$, where $\vec{z}_\perp = (z^2,z^3)$ is the
impact--parameter distance between the two loops in the transverse plane.
The two Wilson loops are schematically shown in Fig. 1.

It is convenient to consider also
the correlation function of two Euclidean Wilson loops $\widetilde{\cal W}_1$
and $\widetilde{\cal W}_2$ running along two rectangular paths
$\widetilde{\cal C}_1$ and $\widetilde{\cal C}_2$ which follow the following
straight--line trajectories:
\bea
\widetilde{\cal C}_1 &:&
X^{1q}_{E\mu}(\tau) = z_{E\mu} + {p_{1E\mu} \over m} \tau
+ (1-f_1) R_{1E\mu} ~~,~~
X^{1\bar{q}}_{E\mu}(\tau) = z_{E\mu} + {p_{1E\mu} \over m} \tau 
- f_1 R_{1E\mu} ,
\nonumber \\
\widetilde{\cal C}_2 &:&
X^{2q}_{E\mu}(\tau) = {p_{2E\mu} \over m} \tau + (1-f_2) R_{2E\mu} ~~,~~
X^{2\bar{q}}_{E\mu}(\tau) = {p_{2E\mu} \over m} \tau - f_2 R_{2E\mu} ,
\label{trajE}
\eea
and are closed by straight--line paths at proper times $\tau = \pm T$. Here
$R_{1E} = (0,\vec{R}_{1\perp},0)$, $R_{2E} = (0,\vec{R}_{2\perp},0)$ and
$z_E = (0,\vec{z}_\perp,0)$. Moreover, in the Euclidean theory we {\it choose}
the four--vectors $p_{1E}$ and $p_{2E}$ to be:
\be
p_{1E} =
m \left( \sin{\theta \over 2}, \vec{0}_\perp, \cos{\theta \over 2} \right) ,~~~
p_{2E} =
m \left( -\sin{\theta \over 2}, \vec{0}_\perp, \cos{\theta \over 2} \right) ,
\label{p1p2E}
\ee
$\theta$ being the angle formed by the two trajectories $1q$ and $2q$ in
Euclidean four--space, i.e., $p_{1E} \cdot p_{2E} = m^2 \cos\theta$.\\
Let us introduce the following notations for the normalized loop--loop
correlators in the Minkowskian and in the Euclidean theory,
in the presence of a {\it finite} IR cutoff $T$:
\be
{\cal G}_M(\chi,T,\vec{z}_\perp;1,2) \equiv
{ \langle {\cal W}^{(T)}_1 {\cal W}^{(T)}_2 \rangle \over
\langle {\cal W}^{(T)}_1 \rangle
\langle {\cal W}^{(T)}_2 \rangle } ,~~~
{\cal G}_E(\theta,T,\vec{z}_\perp;1,2) \equiv
{ \langle \widetilde{\cal W}^{(T)}_1 \widetilde{\cal W}^{(T)}_2 \rangle_E \over
\langle \widetilde{\cal W}^{(T)}_1 \rangle_E
\langle \widetilde{\cal W}^{(T)}_2 \rangle_E } ,
\label{GM-GE}
\ee
where the arguments ``$1$'' and ``$2$'' in the functions ${\cal G}_M$ and
${\cal G}_E$ stand for ``$\vec{R}_{1\perp}, f_1$'' and
``$\vec{R}_{2\perp}, f_2$'' respectively.
The expectation values $\langle \ldots \rangle_E$ are averages
in the sense of the Euclidean functional integrals.\\
The Minkowskian quantity ${\cal G}_M$ with $\chi\in\mathbb{R}^+$ can be
reconstructed from the corresponding Euclidean quantity ${\cal G}_E$,
with $\theta \in (0,\pi)$, by an analytic continuation in the angular
variables $\theta \to -i\chi$ and in the IR cutoff $T \to iT$, exactly as
in the case of Wilson lines \cite{Meggiolaro02,Meggiolaro97,Meggiolaro98}.
This result \cite{Meggiolaro02,Meggiolaro05} is derived under certain
hypotheses of analyticity in the angular variables and in the IR cutoff $T$.
In particular, one makes the assumption \cite{crossing} that
the function ${\cal G}_E$, as a function of the {\it complex}
variable $\theta$, can be {\it analytically extended} from the real segment
$(0 < \re\theta < \pi, \im\theta = 0)$ to a domain ${\cal D}_E$,
which also includes the negative imaginary axis
$(\re\theta = 0+, \im\theta < 0)$; and, therefore,
the function ${\cal G}_M$, as a function of the {\it complex} variable
$\chi$, can be {\it analytically extended} from the positive
real axis $(\re\chi > 0, \im\chi = 0+)$ to a domain
${\cal D}_M = \{ \chi \in \mathbb{C} ~|~ -i\chi \in {\cal D}_E \}$,
which also includes the imaginary segment $(\re\chi = 0, 0 < \im\chi < \pi)$.
The validity of this assumption is confirmed by explicit calculations in
perturbation theory \cite{Meggiolaro97,Meggiolaro05,BB}.
The domains ${\cal D}_E$ and ${\cal D}_M$ are schematically shown in Figs. 2
and 3.
Denoting with $\overline{\cal G}_M$ and $\overline{\cal G}_E$ such analytic
extensions, we then have the following {\it analytic--continuation relations}
\cite{Meggiolaro05,crossing}:
\bea
\overline{\cal G}_E(\theta,T,\vec{z}_\perp;1,2)
&=& \overline{\cal G}_M (i\theta,-iT,\vec{z}_\perp;1,2) ,
\qquad \forall\theta\in {\cal D}_E ;
\nonumber \\
\overline{\cal G}_M(\chi,T,\vec{z}_\perp;1,2)
&=& \overline{\cal G}_E (-i\chi,iT,\vec{z}_\perp;1,2) ,
\qquad \forall\chi\in {\cal D}_M .
\label{analytic}
\eea
As we have said above, the loop--loop correlation functions (\ref{GM-GE}),
both in the Minkowskian and in the Euclidean theory, are expected to be
IR--{\it finite} quantities, i.e., to have finite limits when $T \to \infty$,
differently from what happens in the case of Wilson lines.
One can then define the following loop--loop correlation functions
with the IR cutoff removed:
\bea
{\cal C}_M(\chi,\vec{z}_\perp;1,2) &\equiv&
\displaystyle\lim_{T \to \infty}
\left[ {\cal G}_M(\chi,T,\vec{z}_\perp;1,2) - 1 \right] , \nonumber \\
{\cal C}_E(\theta,\vec{z}_\perp;1,2)
&\equiv& \displaystyle\lim_{T \to \infty}
\left[ {\cal G}_E(\theta,T,\vec{z}_\perp;1,2) - 1 \right] .
\label{C12}
\eea
It has been proved in Ref. \cite{Meggiolaro05} that, under certain analyticity
conditions in the {\it complex} variable $T$ [conditions which are also
sufficient to make the relations (\ref{analytic}) meaningful], the two
quantities (\ref{C12}), obtained {\it after} the removal of the IR cutoff
($T \to \infty$), are still connected by the usual analytic continuation in
the angular variables only:
\bea
\overline{\cal C}_E(\theta,\vec{z}_\perp;1,2) &=&
\overline{\cal C}_M(i\theta,\vec{z}_\perp;1,2) ,
\qquad \forall\theta\in {\cal D}_E ;
\nonumber \\
\overline{\cal C}_M(\chi,\vec{z}_\perp;1,2) &=&
\overline{\cal C}_E(-i\chi,\vec{z}_\perp;1,2) ,
\qquad \forall\chi\in {\cal D}_M .
\label{final}
\eea
This is a highly non--trivial result, whose general validity is discussed
in Ref. \cite{Meggiolaro05}.
The validity of the relation (\ref{final}) for the loop--loop correlators 
in QCD has been also recently verified in Ref. \cite{BB} by an explicit
calculation up to the order ${\cal O}(g^6)$ in perturbation theory.
However we want to stress that the analytic continuation (\ref{analytic}) or
(\ref{final}) is expected to be an {\it exact} result, i.e., not restricted
to some order in perturbation theory or to some other approximation,
and is valid both for the Abelian and the non--Abelian case.

It has been also recently shown in Ref. \cite{crossing} that the
analytic--continuation relations (\ref{analytic})
allow us to deduce non trivial properties of the Euclidean correlator
$\mathcal{G}_E$ under the exchange $\theta\to \pi -\theta$ and of the
Minkowskian correlator $\mathcal{G}_M$ under the exchange $\chi\to i\pi -\chi$,
the so--called {\it crossing--symmetry relations} for loop--loop correlators:
\footnote{Indeed Eqs. (\ref{crossing}) slightly
generalize the corresponding relations found in Ref. \cite{crossing} for the
special case $f_1 = f_2 = 1/2$. The dependence on the longitudinal--momentum
fractions $f_1$ and $f_2$ in the crossing--symmetry relations is easily
understood, following the method outlined in Ref. \cite{crossing}, by
recognizing that the exchange from a given Wilson loop ${\cal W}$ to the
corresponding {\it antiloop} $\overline{\cal W}$ (obtained by exchanging the
quark and the antiquark trajectories) can be made substituting
$\vec{R}_\perp \to -\vec{R}_\perp$ {\it and} $f \to 1-f$.}
\bea
\lefteqn{
\mathcal{G}_E(\pi-\theta,T,\vec{z}_{\perp};\vec{R}_{1\perp},f_1,
\vec{R}_{2\perp},f_2) }
\label{crossing} \\
& & =\mathcal{G}_E(\theta,T,\vec{z}_{\perp};\vec{R}_{1\perp},f_1,
-\vec{R}_{2\perp},1-f_2)
=\mathcal{G}_E(\theta,T,\vec{z}_{\perp};-\vec{R}_{1\perp},1-f_1,
\vec{R}_{2\perp},f_2) ,
\quad\forall\theta\in\mathbb{R} ;
\nonumber \\
\lefteqn{
\overline{\mathcal{G}}_M(i\pi-\chi,T,\vec{z}_{\perp};\vec{R}_{1\perp},f_1,
\vec{R}_{2\perp},f_2) }
\nonumber \\
& & =\mathcal{G}_M(\chi,T,\vec{z}_{\perp};\vec{R}_{1\perp},f_1,
-\vec{R}_{2\perp},1-f_2)
=\mathcal{G}_M(\chi,T,\vec{z}_{\perp};-\vec{R}_{1\perp},1-f_1,
\vec{R}_{2\perp},f_2) ,
\forall\chi\in\mathbb{R}^+ .
\nonumber
\eea
These two relations are valid for every value of the IR cutoff $T$ and so
completely analogous relations also hold for the loop--loop correlation
functions ${\cal C}_M$ and ${\cal C}_E$ with the IR cutoff removed
($T \to \infty$), defined in Eq. (\ref{C12}):
\bea
\lefteqn{
\mathcal{C}_E(\pi-\theta,\vec{z}_{\perp};\vec{R}_{1\perp},f_1,
\vec{R}_{2\perp},f_2) }
\label{crossingC} \\
& & =\mathcal{C}_E(\theta,\vec{z}_{\perp};\vec{R}_{1\perp},f_1,
-\vec{R}_{2\perp},1-f_2)
=\mathcal{C}_E(\theta,\vec{z}_{\perp};-\vec{R}_{1\perp},1-f_1,
\vec{R}_{2\perp},f_2) ,
\quad\forall\theta\in\mathbb{R} ;
\nonumber \\
\lefteqn{
\overline{\mathcal{C}}_M(i\pi-\chi,\vec{z}_{\perp};\vec{R}_{1\perp},f_1,
\vec{R}_{2\perp},f_2) }
\nonumber \\
& & =\mathcal{C}_M(\chi,\vec{z}_{\perp};\vec{R}_{1\perp},f_1,
-\vec{R}_{2\perp},1-f_2)
=\mathcal{C}_M(\chi,\vec{z}_{\perp};-\vec{R}_{1\perp},1-f_1,
\vec{R}_{2\perp},f_2) ,
\forall\chi\in\mathbb{R}^+ .
\nonumber
\eea

\newsection{How a pomeron--like behaviour can be derived}

\noindent
The relation (\ref{final}) allows the derivation of the {\it loop--loop
scattering amplitude} (\ref{scatt-loop}), which we rewrite as
\be
{\cal M}_{(ll)} (s,t;\vec{R}_{1\perp},f_1,\vec{R}_{2\perp},f_2) = -i~2s~
\widetilde{\cal C}_M (\chi \to +\infty, t;1,2) ,
\label{scatt-loop2}
\ee
$\widetilde{\cal C}_M$ being the two--dimensional Fourier transform of
${\cal C}_M$, with respect to the impact parameter $\vec{z}_\perp$,
at transferred momentum $\vec{q}_\perp$ (with $t = -|\vec{q}_\perp|^2$), i.e.,
\be
\widetilde{\cal C}_M (\chi, t;1,2) \equiv
\displaystyle\int d^2 \vec{z}_\perp e^{i \vec{q}_\perp \cdot \vec{z}_\perp}
{\cal C}_M (\chi,\vec{z}_\perp;1,2) ,
\label{CMtilde}
\ee
from the analytic continuation $\theta \to -i\chi$ of the corresponding
Euclidean quantity:
\be
\widetilde{\cal C}_E (\theta, t;1,2) \equiv
\displaystyle\int d^2 \vec{z}_\perp e^{i \vec{q}_\perp \cdot \vec{z}_\perp}
{\cal C}_E (\theta,\vec{z}_\perp;1,2) ,
\label{CEtilde}
\ee
which can be evaluated non--perturbatively by well--known and well--established
techniques available in the Euclidean theory.\\
We remind the reader that the {\it hadron--hadron elastic scattering amplitude}
${\cal M}_{(hh)}$ can be obtained by averaging the loop--loop scattering
amplitude (\ref{scatt-loop2}) over all possible dipole transverse separations
$\vec{R}_{1\perp}$ and $\vec{R}_{2\perp}$ and longitudinal--momentum fractions
$f_1$ and $f_2$ with two proper squared hadron wave functions:
\bea
{\cal M}_{(hh)}(s,t) &=&
\displaystyle\int d^2\vec{R}_{1\perp} \int_0^1 df_1~
|\psi_1(\vec{R}_{1\perp},f_1)|^2
\displaystyle\int d^2\vec{R}_{2\perp} \int_0^1 df_2~
|\psi_2(\vec{R}_{2\perp},f_2)|^2
\nonumber \\
&\times& {\cal M}_{(ll)} (s,t;\vec{R}_{1\perp},f_1,\vec{R}_{2\perp},f_2) .
\label{scatt-hadron}
\eea
(For a detailed description of the procedure leading from the loop--loop
scattering amplitude ${\cal M}_{(ll)}$ to the hadron--hadron elastic
scattering amplitude ${\cal M}_{(hh)}$ we refer the reader to Refs.
\cite{DFK,Nachtmann97,BN,Dosch,pomeron-book}. See also Ref. \cite{LLCM1}
and references therein.)\\
Denoting with ${\cal C}_M^{(hh)}$ and ${\cal C}_E^{(hh)}$ the quantities
obtained by averaging the corresponding loop--loop correlation functions
${\cal C}_M$ and ${\cal C}_E$ over all possible dipole transverse separations
$\vec{R}_{1\perp}$ and $\vec{R}_{2\perp}$ and longitudinal--momentum fractions
$f_1$ and $f_2$, in the same sense as in Eq. (\ref{scatt-hadron}), i.e.,
\bea
{\cal C}_M^{(hh)} (\chi,\vec{z}_\perp) &\equiv&
\displaystyle\int d^2\vec{R}_{1\perp} \int_0^1 df_1~
|\psi_1(\vec{R}_{1\perp},f_1)|^2
\displaystyle\int d^2\vec{R}_{2\perp} \int_0^1 df_2~
|\psi_2(\vec{R}_{2\perp},f_2)|^2
\nonumber \\
&\times& {\cal C}_M (\chi,\vec{z}_\perp;1,2) ,
\nonumber \\
{\cal C}_E^{(hh)} (\theta,\vec{z}_\perp) &\equiv&
\displaystyle\int d^2\vec{R}_{1\perp} \int_0^1 df_1~
|\psi_1(\vec{R}_{1\perp},f_1)|^2
\displaystyle\int d^2\vec{R}_{2\perp} \int_0^1 df_2~
|\psi_2(\vec{R}_{2\perp},f_2)|^2
\nonumber \\
&\times& {\cal C}_E (\theta,\vec{z}_\perp;1,2) ,
\label{CMEhh}
\eea
we can write:
\be
{\cal M}_{(hh)} (s,t) = -i~2s~
\widetilde{\cal C}_M^{(hh)} (\chi \to +\infty, t) ,
\label{scatt-hadron2}
\ee
where, as usual:
\be
\widetilde{\cal C}_M^{(hh)} (\chi, t) \equiv
\displaystyle\int d^2 \vec{z}_\perp e^{i \vec{q}_\perp \cdot \vec{z}_\perp}
{\cal C}_M^{(hh)} (\chi;\vec{z}_\perp) ,~~~
\widetilde{\cal C}_E^{(hh)} (\theta, t) \equiv
\displaystyle\int d^2 \vec{z}_\perp e^{i \vec{q}_\perp \cdot \vec{z}_\perp}
{\cal C}_E^{(hh)} (\theta;\vec{z}_\perp) .
\label{CMEhh-tilde}
\ee
Clearly, by virtue of the relation (\ref{final}), we also have that:
\be
\overline{\widetilde{\cal C}_M^{(hh)}} (\chi, t) =
\overline{\widetilde{\cal C}_E^{(hh)}} (-i\chi, t) ,
\qquad \forall\chi\in {\cal D}_M .
\label{final-hh}
\ee
We also remind the reader that, in order to obtain the correct
$s$--dependence of the scattering amplitude (\ref{scatt-hadron2}), one must
express the hyperbolic angle $\chi$ between the two loops in terms of $s$,
in the high--energy limit $s \to \infty$ (i.e., $\chi \to +\infty$):
\be
s \equiv (p_1 + p_2)^2 = 2m^2 \left( \cosh\chi + 1 \right) ~,~~~~{\rm i.e.:}~~
\chi \mathop{\sim}_{s \to \infty} \log \left( {s \over m^2} \right) ,
\label{logs}
\ee
where $m$ is the mass of the two hadrons considered.\\
This approach has been extensively used in the literature in order to
tackle, from a theoretical point of view, the still unsolved problem
of the asymptotic $s$--dependence of hadron--hadron elastic scattering
amplitudes and total cross sections.\\
For example, in Ref. \cite{LLCM2} the loop--loop Euclidean correlation
functions have been evaluated in the context of the so--called {\it loop--loop
correlation model} \cite{LLCM1}, in which the QCD vacuum is described by
perturbative gluon exchange and the non--perturbative {\it Stochastic Vacuum
Model} (SVM), and then they have been continued to the corresponding
Minkowskian correlation functions using the above--mentioned analytic
continuation in the angular variables: the result is an $s$--independent
correlation function
$\widetilde{\cal C}_M (\chi \to +\infty, t;1,2)$
and, therefore, a loop--loop scattering amplitude (\ref{scatt-loop2})
linearly rising with $s$. By virtue of the {\it optical theorem},
\be
\sigma_{\rm tot}^{(hh)} (s) \mathop{\sim}_{s \to \infty}
{1 \over s} {\rm Im} {\cal M}_{(hh)} (s, t=0) ,
\label{optical}
\ee
this should imply (apart from possible $s$--dependences in the hadron wave
functions!) $s$--independent hadron--hadron total cross sections in the
asymptotic high--energy limit, in apparent contradiction to the experimental
observations, which seem to be well described by a {\it pomeron}--like
high--energy behaviour (see, for example, Ref. \cite{pomeron-book} and
references therein):
\be
\sigma_{\rm tot}^{(hh)} (s) \mathop{\sim}_{s \to \infty}
\sigma_0^{(hh)} \left( {s \over s_0} \right)^{\epsilon_P} ,
~~~~{\rm with:}~~\epsilon_P \simeq 0.08 .
\label{pomeron}
\ee
In Refs. \cite{DFK,BN} a possible $s$--dependence in the hadron wave
functions was advocated in order to reproduce the phenomenological
{\it pomeron}--like high--energy behaviour of the total cross sections.
However, it would be surely preferable to ascribe the {\it universal}
high--energy behaviour of hadron--hadron total cross sections [the only
dependence on the initial--state hadrons being in the multiplicative constant
$\sigma_0^{(hh)}$ in Eq. (\ref{pomeron})] to the same {\it fundamental}
quantity, i.e., the loop--loop scattering amplitude.
(For a different, but still phenomenological, approach in this direction,
using the SVM, see Ref. \cite{LLCM1}.)\\
The same approach, based on the analytic continuation from Euclidean to
Minkowskian correlation functions, has been also adopted in Ref.
\cite{instanton1} in order to study the one--instanton contribution to
both the line--line (see also Ref. \cite{instanton2}) and the loop--loop
scattering amplitudes: one finds that, after the analytic continuation,
the colour--elastic line--line and loop--loop correlation functions decay
as $1/s$ with the energy. (Instead, the colour--changing inelastic line--line
correlation function is of order $s^0$ and dominates at high energy.
In a further paper \cite{instanton3}, instanton--induced inelastic collisions
have been investigated in more detail and shown to produce total cross sections
increasing as $\log s$.)\\
A behaviour like the one of Eq. (\ref{pomeron}) seems to emerge directly
(apart from possible undetermined $\log s$ prefactors) when applying
the Euclidean--to--Minkowskian analytic--continuation approach to the
study of the line--line/loop--loop scattering amplitudes in strongly coupled
(confining) gauge theories using the AdS/CFT correspondence \cite{JP2,Janik}.
(In a previous paper \cite{JP1} the same approach was also used to study
the loop--loop scattering amplitudes in the ${\cal N} = 4$ SYM theory in
the limit of large number of colours, $N_c \to \infty$, and strong coupling.)

As we have already remembered in the previous section, after Eq. (\ref{final}),
the loop--loop correlation functions (both in the Minkowskian and in the
Euclidean theory) have been computed exactly in the first two orders of
perturbation theory, ${\cal O}(g^4)$ and ${\cal O}(g^6)$, in Ref. \cite{BB}.
(Strictly speaking, the loop--loop correlators are considered in Ref. \cite{BB}
in a different context, as {\it elementary} high--energy scattering processes
used to reconstruct, after proper integration over dipole parameters and
separations, the high--energy scattering amplitude of two {\it virtual
photons}, where each photon splits into a quark--antiquark dipole.)
There are two basic results in Ref. \cite{BB}: the first result is that the
loop--loop correlation function is an analytic function of the angle between
the dipoles, so confirming Eq. (\ref{final}). The second basic result is that
the dipole--dipole cross section, evaluated from the loop--loop correlator
up to the order ${\cal O}(g^6)$, reproduces the first iteration of the BFKL
{\it kernel} in the leading--log approximation, the so--called
BFKL--{\it pomeron} behaviour, i.e., $\sim s^{{12\alpha_s \over \pi}\log 2}$,
with $\alpha_s = g^2/4\pi$ \cite{BFKL}. (Even if the authors of Ref. \cite{BB}
have no access to the next--to--leading--order BFKL terms, since this would
require the formidable computation of the loop--loop correlators up to the order
${\cal O}(g^8)$, still they conclude [and we agree!] by saying that, by virtue
of the analyticity of the loop--loop correlation function in the angle, in
principle one can get the {\it full} BFKL {\it kernel} from an Euclidean
calculation.)

The way in which a {\it pomeron}--like behaviour can emerge, using the
Euclidean--to--Minkowskian analytic continuation, was first shown in Ref.
\cite{Meggiolaro97} in the case of the line--line (i.e., parton--parton)
scattering amplitudes. Here we shall readapt that analysis to the case of the
loop--loop scattering amplitudes, with more technical developments, new
interesting insights and critical considerations.\\
We start by writing the Euclidean hadronic correlation function
in a partial--wave expansion:
\be
\widetilde{\cal C}_E^{(hh)} (\theta,t) =
\displaystyle\sum_{l=0}^{\infty} (2l+1) A_l(t) P_l (\cos \theta) ,
\label{pwe}
\ee
which, by virtue of the orthogonality relation of the Legendre polynomials:
\be
\displaystyle\int_{-1}^{+1} dz~ P_l(z) P_{l'}(z) = {2 \over 2l+1} \delta_{ll'} ,
\ee
can be inverted to give the partial--wave {\it amplitudes}:
\be
A_l(t) = {1 \over 2} \displaystyle\int_{-1}^{+1} d\cos\theta~ P_l(\cos\theta)~
\widetilde{\cal C}_E^{(hh)} (\theta,t) .
\label{pw-ampl}
\ee
As shown in Ref. \cite{crossing} (and briefly recalled at the end of the
previous section), the loop--antiloop correlator at angle $\theta$ in the
Euclidean theory (or at hyperbolic angle $\chi$ in the Minkowskian theory) can
be derived from the corresponding loop--loop correlator by the substitution
$\theta \to \pi - \theta$ (or $\chi \to i\pi - \chi$ in the Minkowskian
theory). Because of these {\it crossing--symmetry relations}, it is natural
to decompose also our hadronic correlation function
$\widetilde{\cal C}_E^{(hh)} (\theta,t)$ as a sum of a
{\it crossing--symmetric} function $\widetilde{\cal C}_E^+ (\theta,t)$
and of a {\it crossing--antisymmetric} function
$\widetilde{\cal C}_E^- (\theta,t)$:
\bea
\widetilde{\cal C}_E^{(hh)} (\theta,t) &=&
\widetilde{\cal C}_E^+ (\theta,t) + \widetilde{\cal C}_E^- (\theta,t) ,
\nonumber \\
\widetilde{\cal C}_E^{\pm} (\theta,t) &\equiv&
{\widetilde{\cal C}_E^{(hh)} (\theta,t) \pm \widetilde{\cal C}_E^{(hh)}
(\pi-\theta,t) \over 2} .
\eea
Using Eq. (\ref{pwe}), we can find the partial--wave expansions of these two
functions as follows:
\be
\widetilde{\cal C}_E^{\pm} (\theta,t) = {1 \over 2}
\displaystyle\sum_{l=0}^{\infty} (2l+1) A_l(t)
[P_l (\cos \theta) \pm P_l (-\cos \theta)] .
\label{pwe-pm}
\ee
Because of the relation $P_l(-\cos \theta) = (-1)^l P_l(\cos \theta)$, valid
for non--negative integer values of $l$, we immediately see that
$\widetilde{\cal C}_E^+ (\theta,t)$ gets contributions only from even
$l$, while $\widetilde{\cal C}_E^- (\theta,t)$ gets contributions only from
odd $l$. For this reason the functions
$\widetilde{\cal C}_E^{\pm} (\theta,t)$ can also be called
{\it even--signatured} and {\it odd--signatured} correlation functions
respectively and we can replace $A_l(t)$ in Eq. (\ref{pwe-pm}) respectively
with $A_l^\pm (t) \equiv {1 \over 2} [1 \pm (-1)^l] A_l(t)$, that is:
\be
A_l^+ (t) = \left\{
\begin{matrix}
A_l(t) &,& {\rm for~ even}~ l \\
0 &,& {\rm for~ odd}~ l
\end{matrix}
\right. ;~~~
A_l^- (t) = \left\{
\begin{matrix}
0 &,& {\rm for~ even}~ l \\
A_l(t) &,& {\rm for~ odd}~ l
\end{matrix}
\right. .
\label{constraint}
\ee
However, if we write the hadronic correlation function
$\widetilde{\cal C}_E^{(hh)} (\theta,t)$, by virtue of Eqs. (\ref{CMEhh-tilde})
and (\ref{CMEhh}), in terms of the loop--loop correlation function, averaged
over all possible dipole transverse separations $\vec{R}_{1\perp}$ and
$\vec{R}_{2\perp}$ and longitudinal--momentum fractions $f_1$ and $f_2$ with
two proper squared hadron wave functions 
$|\psi_1(\vec{R}_{1\perp},f_1)|^2$ and $|\psi_2(\vec{R}_{2\perp},f_2)|^2$,
and we make use:
i) of the crossing--symmetry relations (\ref{crossingC}), and ii) of the
rotational-- and $C$--invariance of the squared hadron wave functions, that is
$|\psi_i(\vec{R}_{i\perp},f_i)|^2 = |\psi_i(-\vec{R}_{i\perp},f_i)|^2 =
|\psi_i(\vec{R}_{i\perp},1-f_i)|^2 = |\psi_i(-\vec{R}_{i\perp},1-f_i)|^2$
(see Refs. \cite{Dosch,LLCM1} and also \cite{pomeron-book}, chapter 8.6, and
references therein), then we immediately conclude that the hadronic correlation
function $\widetilde{\cal C}_E^{(hh)} (\theta,t)$ is automatically
crossing symmetric and so it coincides with the even--signatured
function $\widetilde{\cal C}_E^+ (\theta,t)$, the odd--signatured
function $\widetilde{\cal C}_E^- (\theta,t)$ being identically equal to zero.
Upon analytic continuation from the Euclidean to the Minkowskian theory, this
means that the Minkowskian hadronic correlation function
$\widetilde{\cal C}_M^{(hh)} (\chi,t)$, and therefore also the scattering
amplitude ${\cal M}_{(hh)}$ written in Eq. (\ref{scatt-hadron2}), turns out to
be automatically crossing symmetric, i.e., invariant under the exchange
$\chi \to i\pi - \chi$:
$\widetilde{\cal C}_M^{(hh)} (\chi,t) = \widetilde{\cal C}_M^+ (\chi,t)$,
$\widetilde{\cal C}_M^- (\chi,t) = 0$.
In other words, our formalism naturally leads to a high--energy meson--meson
scattering amplitude which, being crossing symmetric, automatically
satisfies the Pomeranchuk theorem. An {\it odderon} (i.e., $C=-1$) exchange
seems to be excluded for high--energy meson--meson scattering, while a
{\it pomeron} (i.e., $C=+1$) exchange is possible.
(This conclusion about the {\it odderon} suppression in meson--meson
scattering agrees with that of Ref. \cite{odderon}. It would be interesting
to see if and how this conclusion would change in a more general context,
i.e., by generalizing our approach, based on the Euclidean--to--Minkowskian
analytic continuation, to the case in which baryons and antibaryons are
involved. This can surely be done, but we shall not tackle this problem in
the present paper, where we are mainly interested in the {\it pomeron}, and
we prefer to leave it to a future publication.)

Let us therefore proceed by considering our {\it crossing--symmetric}
Euclidean correlation function:
\be
\widetilde{\cal C}_E^{(hh)} (\theta,t) =
\widetilde{\cal C}_E^+ (\theta,t) = {1 \over 2}
\displaystyle\sum_{l=0}^{\infty} (2l+1) A_l^+ (t)
[P_l (\cos \theta) + P_l (-\cos \theta)] .
\label{pwe-E}
\ee
We can now use Cauchy's theorem to rewrite this partial--wave expansion as an
integral over $l$, the so--called {\it Sommerfeld--Watson transform}:
\be
\widetilde{\cal C}_E^{(hh)} (\theta,t) = \widetilde{\cal C}_E^+ (\theta,t) =
-{1 \over 4i} \displaystyle\int_C {(2l+1) A_l^+(t)
[ P_l(-\cos\theta) + P_l(\cos\theta)] \over \sin(\pi l)} dl ,
\label{swt}
\ee
where ``$C$'' is a contour in the complex $l$--plane, running clockwise
around the real positive $l$--axis and enclosing all non--negative integers,
while excluding all the singularities of $A_l$. (Eq. (\ref{swt}) can be
verified after recognizing that $P_l (\pm\cos \theta)$ is an integer function
of $l$ and that the singularities enclosed by the contour $C$ of the expression
under integration in the Eq. (\ref{swt}) are only the simple poles of
$1/\sin(\pi l)$ at the non--negative integer values of $l$.)
Here (as in the original derivation! But see below for more comments about
the comparison between our approach and the original one) we make the
fundamental {\it assumption} that the singularities of $A_l(t)$ in the complex
$l$--plane (at a given $t$) are only {\it simple poles}. Then we can use again
Cauchy's theorem to reshape the contour $C$ into the straight line
$\re (l) = -{1 \over 2}$ and rewrite the integral (\ref{swt}) as follows:
\bea
\lefteqn{
\widetilde{\cal C}_E^{(hh)} (\theta,t) = \widetilde{\cal C}_E^+ (\theta,t) = }
\nonumber \\
& & -{\pi \over 2} \displaystyle\sum_{ \re (\sigma_n^+) >  -{1 \over 2} }
{ (2\sigma_n^+ (t) + 1) r_n^+ (t) [P_{\sigma_n^+ (t)} (-\cos \theta)
+ P_{\sigma_n^+ (t)}(\cos \theta)] \over \sin (\pi \sigma_n^+ (t))}
\nonumber \\
& & -{1 \over 4i}
\displaystyle\int_{-{1 \over 2}-i\infty}^{-{1 \over 2}+i\infty}
{ (2l+1) A_l^+ (t) [P_l (-\cos \theta) + P_l (\cos \theta)] \over
\sin (\pi l) } dl ,
\label{swt-E}
\eea
where $\sigma_n^+ (t)$ is a pole of $A_l^+ (t)$ in the complex $l$--plane and
$r_n^+ (t)$ is the corresponding residue. We have also assumed that the
large--$l$ behaviour of $A_l^+$ is such that the integrand function in
Eq. (\ref{swt}) vanishes enough rapidly (faster than
$1/l$) as $|l| \to \infty$ in the right half--plane, so that the 
contribution from the infinite contour is zero.
As it is shown in the Appendix A of Ref. \cite{pomeron-book}, a necessary
condition, in order to satisfy this requirement on the large--$l$ behaviour,
is that $A_l^+ (t)$ does not diverge faster than $e^{{\pi \over 2} |l|}$ for
large $l$. A theorem, known as {\it Carlson's theorem} (see, e.g., Ref.
\cite{Titchmarsh}, p.186), then ensures that $A_l^+ (t)$, because of the
above--mentioned requirement, is defined uniquely: we cannot add a (non--zero)
term to $A_l^+ (t)$ which at the same time preserves the constraint
(\ref{constraint}), while maintaining the required asymptotic behaviour.
(In the original derivation of the {\it Regge poles} [see, e.g., Refs.
\cite{pomeron-book} and \cite{Collins}], one can find a proper definition
of the partial--wave amplitudes $A_l^{\pm}$ in the complex $l$--plane
by using the so--called {\it Froissart--Gribov formula}, that satisfies the
constraints (\ref{constraint}) at physical, i.e., non--negative integer, values
of $l$ and vanishes exponentially for large $l$: then as we have commented
above, {\it Carlson's theorem} ensures that this definition is unique.
In principle one can try to follow a similar approach also in our case, by
rewriting Eq. (\ref{pw-ampl}), defining the partial--wave amplitudes,
expressing the Legendre function $P_l(\cos \theta)$ in terms of the Legendre
functions of the second kind $Q_l$ [see Ref. \cite{GR}, relations {\bf 8.820}
9 and {\bf 8.834} 1]:
\be
A_l(t) = {i \over 2\pi} \displaystyle\int_{-1}^{+1} d\cos\theta~
[Q_l(\cos\theta + i\varepsilon) - Q_l(\cos\theta - i\varepsilon)]
~\widetilde{\cal C}_E^{(hh)} (\theta,t) , ~~~ {\rm with}~ \varepsilon \to 0+ .
\ee
However, in order to go on with the technical manipulations [see, e.g., Ref.
\cite{pomeron-book}, par. 1.6 and Appendix A] that lead to the
{\it Froissart--Gribov formula}, or at least to some equivalent new version
of it, we need to make some nontrivial assumptions about the nature (type
and location) of the singularities of the Euclidean correlation function
$\widetilde{\cal C}_E^{(hh)} (\theta,t)$ in the complex $\theta$--plane.
Unfortunately, as it has also been recently well remarked in Ref.
\cite{crossing}, too little is known with regard to this problem: one should
find a nonperturbative way of identifying all type of singularities in the
correlators and so have a complete description of their analyticity structure.
We do not tackle this interesting and formidable problem in this paper
[leaving it to future works] and we content ourselves in {\it assuming} the
existence of such a function $A_l^+ (t)$, defined in the complex $l$--plane,
satisfying the constraint (\ref{constraint}) at physical, i.e., non--negative
integer, values of $l$ and the above--mentioned requirement on the asymptotic
large--$l$ behaviour.)

Eq. (\ref{swt-E}) immediately
leads to the asymptotic behaviour of the scattering amplitude in the limit
$s \to \infty$, with a fixed $t$ ($|t| \ll s$). In fact, making use of the
analytic extension (\ref{final-hh}) when continuing the angular variable,
$\theta \to -i\chi$, we derive that for every $\chi\in\mathbb{R}^+$:
\bea
\lefteqn{
\widetilde{\cal C}_M^{(hh)} (\chi, t) =
\overline{\widetilde{\cal C}_E^{(hh)}} (-i\chi, t) = } \nonumber \\
& & -{\pi \over 2} \displaystyle\sum_{ \re (\sigma_n^+) >  -{1 \over 2} }
{ (2\sigma_n^+ (t) + 1) r_n^+ (t) [P_{\sigma_n^+ (t)} (-\cosh \chi)
+ P_{\sigma_n^+ (t)}(\cosh \chi)] \over \sin (\pi \sigma_n^+ (t))}
\nonumber \\
& & -{1 \over 4i}
\displaystyle\int_{-{1 \over 2}-i\infty}^{-{1 \over 2}+i\infty}
{ (2l+1) A_l^+ (t) [P_l (-\cosh \chi) + P_l (\cosh \chi)] \over
\sin (\pi l) } dl ,
\label{swt-M}
\eea
Now we must take the large--$\chi$ (large--$s$) limit of this expression, with
the hyperbolic angle $\chi$ expressed in terms of $s$ by the relation
(\ref{logs}), i.e., $\cosh \chi = {s \over 2m^2} - 1$.
The asymptotic form of $P_\nu (z)$ when $z \to \infty$ is known to be
a linear combination of $z^\nu$ and of $z^{-\nu -1}$
(see Ref. \cite{GR}, relation {\bf 8.776} 1):
\be
P_\nu (z) \mathop\sim_{z \to \infty} {1 \over \sqrt{\pi}}
\left[ {\Gamma (\nu + {1 \over 2}) \over \Gamma (\nu + 1)} (2z)^\nu
+ {\Gamma (-\nu - {1 \over 2}) \over \Gamma (-\nu)} (2z)^{-\nu-1} \right] .
\ee
When $\re (\nu) >  -1/2$, the last term can be neglected and thus we obtain,
for each term in the sum in Eq. (\ref{swt-M}):
\be
P_{\sigma_n^+} (\cosh \chi) + P_{\sigma_n^+} (-\cosh \chi)
\mathop\sim_{s \to \infty}
\left[ 1 + e^{-i\pi\sigma_n^+} \right] {1 \over \sqrt{\pi}}
{\Gamma (\sigma_n^+ + {1 \over 2}) \over \Gamma (\sigma_n^+ + 1)}
\left( {s \over m^2} \right)^{\sigma_n^+} ,
\ee
where for $P_{\sigma_n^+} (-\cosh \chi)$ we have used the relation
{\bf 8.776} 2 of Ref. \cite{GR}:
\be
P_\nu (-z) = e^{-i\pi\nu} P_\nu (z) - {2 \over \pi} \sin (\pi\nu) Q_\nu (z),
~~~ {\rm for}~~\im (z) > 0 ,
\label{Pcut}
\ee
with the following large--$z$ behaviour of the Legendre functions of the second
kind $Q_\nu (z)$ (see Ref. \cite{GR}, relation {\bf 8.776} 2):
\be
Q_\nu (z) \mathop\sim_{z \to \infty} \sqrt{\pi}
{\Gamma (\nu + 1) \over \Gamma (\nu + {3 \over 2})} (2z)^{-\nu-1} .
\ee
Let us observe that we have used Eq. (\ref{Pcut}), valid for $\im(z)>0$, since
in our case $z = \cosh \chi = {s \over 2m^2} - 1$: if (following the usual
$i\varepsilon$--{\it prescription} used both in perturbation theory and also
outside the framework of perturbation theory) we provide the squared mass
$m^2$ with a small negative imaginary part, i.e., $m^2 \to m^2 - i\varepsilon$,
with $\varepsilon \to 0+$, then $z$ acquires a small positive imaginary part.
In other words, the physical ($s$--channel) scattering amplitude is reached
by analytic continuation in $s$ down on to the positive real axis from the
upper half of the complex $s$--plane, i.e., $s \to s + i\varepsilon$, with
$\varepsilon \to 0+$, as is well known.
Or, equivalently, in our formalism based on the analytic continuation of the
loop--loop correlators in the angular variables, the physical ($s$--channel)
scattering amplitude is obtained by analytic continuation of
$(-i2s) \widetilde{\cal C}_M^{(hh)} (\chi, t)$ in the variable $\chi$ down on
to the positive real axis from the upper half of the complex $\chi$--plane,
i.e., $\chi \to \chi + i\varepsilon$, with $\varepsilon \to 0+$; that is to say
[by virtue of Eqs. (\ref{final-hh})], by analytic continuation of
$(-i2s) \widetilde{\cal C}_E^{(hh)} (\theta, t)$ in the variable $\theta$ down
on to the negative imaginary axis from the right--hand half of the complex
$\theta$--plane, i.e., $\theta \to -i\chi + \varepsilon = -i(\chi +
i\varepsilon)$, with $\varepsilon \to 0+$.

Therefore, in the limit $s \to \infty$, with a fixed $t$ ($|t| \ll s$),
we are left with the following expression:
\be
\widetilde{\cal C}_M^{(hh)} \left( \chi \mathop{\sim}_{s \to \infty}
\log \left( {s \over m^2} \right), t \right)
\sim \displaystyle\sum_{ \re (\sigma_n^+) >  -{1 \over 2} }
\beta_n^+ (t) s^{\sigma_n^+ (t)} .
\label{CMhh-regge}
\ee
The integral in Eq. (\ref{swt-M}), usually called the {\it background term},
vanishes at least as $1/\sqrt{s}$.
Eq. (\ref{CMhh-regge}) allows to immediately extract the scattering
amplitude according to Eq. (\ref{scatt-hadron2}):
\be
{\cal M}_{(hh)} (s,t) \mathop{\sim}_{s \to \infty}
-2i \displaystyle\sum_{ \re (\sigma_n^+) > -{1 \over 2} }
\beta_n^+ (t) s^{1+\sigma_n^+ (t)} .
\label{Mhh-regge}
\ee
This equation gives the explicit $s$--dependence of the scattering amplitude at 
very high energy ($s \to \infty$) and small transferred momentum ($|t| \ll s$).
As we can see, this amplitude comes out to be a sum of powers of $s$.
This sort of behaviour for the scattering amplitude is known in the literature
as a {\it Regge behaviour} and $1+\sigma_n^+ (t) \equiv \alpha_n^+ (t)$ is
the so--called {\it Regge trajectory}. In the original derivation
(see, e.g., Refs. \cite{pomeron-book} and \cite{Collins}) the asymptotic
behaviour (\ref{Mhh-regge}) is recovered by analytically continuing the
$t$--channel scattering amplitude to very large imaginary values of the angle
between the trajectories of the two exiting particles in the $t$--channel
scattering process.
Instead, in our derivation, we have used the Euclidean--to--Minkowskian
analytic continuation (\ref{final-hh}) and we have analytically continued
the Euclidean loop--loop correlator to very large (negative) imaginary values 
of the angle $\theta$ between the two Euclidean Wilson loops.
As in the original derivation, we have assumed that the singularities of
$A_l^+ (t)$ in the complex $l$--plane (at a given $t$) are only simple poles
in $l = \sigma_n^+ (t)$: in the original approach, these are known as
{\it Regge poles}, so named after the seminal papers by Regge
in the framework of non--relativistic potential scattering \cite{Regge}.
However, we want to remark that our partial--wave {\it amplitudes} $A_l^+ (t)$
are {\it not} the same partial--wave amplitudes considered in the original
derivation; and, while in the original derivation each {\it Regge pole}
in $l = \sigma (t)$ contributes to the scattering amplitude ${\cal M}$ with a
term of the type $\sim s^{\sigma (t)}$, in our approach the exponent in the
contribution to the scattering amplitude $\sim s^{\alpha (t)}$ differs by 1
from the corresponding pole of $A_l^+ (t)$, i.e.,
$\alpha (t) = 1 + \sigma (t)$, as shown in Eq. (\ref{Mhh-regge}).
Of course, if there are other kinds of singularities, different from simple
poles, their contribution will be of a different type and, in general, also
logarithmic terms (of $s$) may appear in the amplitude. (For example, a triple
pole in $l = \sigma (t)$ would give rise to a contribution in the amplitude
of the type $\sim s^{1 + \sigma (t)} (\log s)^2$, that is, by virtue of the
optical theorem (\ref{optical}), to a contribution in the cross section of
the type $\sim s^{\re[\sigma (0)]} (\log s)^2$, which for $\re[\sigma (0)] = 0$
has exactly the form of the {\it Froissart bound} in Eq. (\ref{FLM}), that
we shall discuss below.)

Denoting with $\sigma_P (t)$ the pole with the largest real part (at
that given $t$) and with $\beta_P (t)$ the corresponding coefficient
$\beta_n^+ (t)$ in Eq. (\ref{CMhh-regge}), we thus find that:
\be
\widetilde{\cal C}_M^{(hh)} \left( \chi \mathop{\sim}_{s \to \infty}
\log \left( {s \over m^2} \right), t \right)
\sim \beta_P (t) s^{\sigma_P (t)} .
\label{CMhh-asympt}
\ee
This implies, for the hadron--hadron elastic scattering amplitude
(\ref{Mhh-regge}), the following high--energy behaviour:
\be
{\cal M}_{(hh)} (s,t) \mathop{\sim}_{s \to \infty}
-2i~\beta_P (t)~ s^{\alpha_P (t)} ,
\label{Mhh-asympt}
\ee
where $\alpha_P (t) \equiv 1 + \sigma_P (t)$ is the {\it pomeron trajectory}.
Therefore, by virtue of the optical theorem (\ref{optical}):
\be
\sigma_{\rm tot}^{(hh)} (s) \mathop{\sim}_{s \to \infty}
\sigma_0^{(hh)} \left( {s \over s_0} \right)^{\epsilon_P} ,
~~~~{\rm with:}~~\epsilon_P = \re[\alpha_P (0)]-1 .
\label{pomeron2}
\ee
We want to stress two important issues which clarify under which conditions
we have been able to derive this {\it pomeron}--like behaviour for the
elastic amplitudes and the total cross sections.

{\bf i)} We have ignored a possible energy dependence of hadron wave functions
and we have thus ascribed the high--energy behaviour of the Minkowskian
hadronic correlation function exclusively to the {\it fundamental} loop--loop
correlation function (\ref{CMtilde}). With this hypothesis, the coefficients
$A_l^+$ in the partial--wave expansion (\ref{pwe}) and, as a consequence, the
coefficients $\beta_n^+$ and $\sigma_n^+$ in the Regge expansion
(\ref{CMhh-regge}) do not depend on $s$, but they only depend on the
Mandelstam variable $t$.

{\bf ii)} However, this is not enough to guarantee the experimentally--observed
{\it universality} (i.e., independence on the specific type of hadrons
involved in the reaction) of the {\it pomeron} trajectory
$\alpha_P (t)$ in Eq. (\ref{Mhh-asympt}) and, therefore, of the
{\it pomeron} intercept $1+\epsilon_P$ in Eq. (\ref{pomeron2}).
In fact, the partial--wave expansion (\ref{pwe})
of the hadronic correlation function can be considered, by virtue of
Eqs. (\ref{CMEhh}) and (\ref{CMEhh-tilde}), as a result of a partial--wave
expansion of the loop--loop Euclidean correlation function
(\ref{CEtilde}), i.e.,
\be
\widetilde{\cal C}_E (\theta,t;1,2) =
\displaystyle\sum_{l=0}^{\infty} (2l+1) {\cal A}_l(t;1,2) P_l (\cos \theta) ,
\label{CEtilde-pwe}
\ee
which is then averaged with two proper squared hadron wave functions:
\bea
\widetilde{\cal C}_E^{(hh)} (\theta,t) &=&
\displaystyle\int d^2\vec{R}_{1\perp} \int_0^1 df_1~
|\psi_1(\vec{R}_{1\perp},f_1)|^2
\displaystyle\int d^2\vec{R}_{2\perp} \int_0^1 df_2~
|\psi_2(\vec{R}_{2\perp},f_2)|^2
\nonumber \\
&\times& \widetilde{\cal C}_E (\theta,t;1,2) .
\label{CEhh-tilde}
\eea
If we now repeat for the partial--wave expansion (\ref{CEtilde-pwe}) the
same manipulations that have led us from Eq. (\ref{pwe}) to Eq.
(\ref{CMhh-regge}), we arrive at the following Regge expansion for the
(even--signatured) loop--loop Minkowskian correlator:
\be
\widetilde{\cal C}_M^+ \left( \chi \mathop{\sim}_{s \to \infty}
\log \left( {s \over m^2} \right), t;1,2 \right)
\sim \displaystyle\sum_{ \re (a_n^+) >  -{1 \over 2} }
b_n^+ (t;1,2) s^{a_n^+ (t;1,2)} ,
\label{CM-regge}
\ee
where $a_n^+ (t;1,2)$ is a pole of ${\cal A}_l^+ (t;1,2)$ in the complex
$l$--plane. After inserting the expansion (\ref{CM-regge}) into the expression
for the Minkowskian hadronic correlation function:
\bea
\widetilde{\cal C}_M^{(hh)} (\chi,t) &=&
\displaystyle\int d^2\vec{R}_{1\perp} \int_0^1 df_1~
|\psi_1(\vec{R}_{1\perp},f_1)|^2
\displaystyle\int d^2\vec{R}_{2\perp} \int_0^1 df_2~
|\psi_2(\vec{R}_{2\perp},f_2)|^2
\nonumber \\
&\times& \widetilde{\cal C}_M^+ (\chi,t;1,2) ,
\label{CMhh-tilde}
\eea
one in general finds a high--energy behaviour which hardly fits with that
reported in Eqs. (\ref{CMhh-asympt}) and (\ref{Mhh-asympt}) with a
universal {\it pomeron} trajectory $\alpha_P (t)$, {\it unless}
one assumes that, for each given loop--loop correlation function with
transverse separations $\vec{R}_{1\perp}$ and $\vec{R}_{2\perp}$ and
longitudinal--momentum fractions $f_1$ and $f_2$, (at least)
the location of the pole $a_n^+ (t;1,2)$ with the largest real part does not
depend on $\vec{R}_{1\perp},f_1$ and $\vec{R}_{2\perp},f_2$,
but only depends on $t$. (Maybe this is a rather natural assumption if
one believes that the {\it pomeron} trajectory is, after all, determined by an
even more fundamental quantity, that is the line--line, i.e., parton--parton,
correlation function.)
If we denote this common pole with $\sigma_P (t)$ and the
corresponding coefficient $b_n^+ (t;1,2)$ in Eq. (\ref{CM-regge}) with
$b_P (t;1,2)$, we then immediately recover the high--energy behaviour
(\ref{CMhh-asympt}), where the coefficient in front is given by:
\be
\beta_P (t) =
\displaystyle\int d^2\vec{R}_{1\perp} \int_0^1 df_1~
|\psi_1(\vec{R}_{1\perp},f_1)|^2
\displaystyle\int d^2\vec{R}_{2\perp} \int_0^1 df_2~
|\psi_2(\vec{R}_{2\perp},f_2)|^2
~b_P (t;1,2) .
\label{beta-tilde}
\ee
This coefficient, differently from the universal function
$\alpha_P (t) = 1 + \sigma_P (t)$, explicitly depends on the specific type of
hadrons involved in the process.

\newsection{Conclusions and outlook}

\noindent
In conclusion, we have shown that the Euclidean--to--Minkowskian
analytic--continuation approach can, with the inclusion of some extra
(more or less plausible) assumptions, easily reproduce a {\it pomeron}--like
behaviour for the high--energy total cross sections, in apparent agreement
with the present--day experimental observations.
However, we should also keep in mind that the {\it pomeron}--like behaviour
(\ref{pomeron}) is, strictly speaking, theoretically forbidden (at least if
considered as a true {\it asymptotic} behaviour) by the well--known
Froissart--Lukaszuk--Martin (FLM) theorem \cite{FLM} (see also
\cite{Heisenberg}), according to which, for $s \to \infty$:
\be
\sigma_{\rm tot}(s) \le {\pi \over m_\pi^2} \log^2 \left( {s \over s_0}
\right) ,
\label{FLM}
\ee
where $m_\pi$ is the pion mass and $s_0$ is an unspecified squared mass
scale.\\
In this respect, the {\it pomeron}--like behaviour (\ref{pomeron}) can at most
be regarded as a sort of {\it pre--asymptotic} (but not really
{\it asymptotic}!) behaviour of the high--energy total cross sections
(see, e.g., Refs. \cite{BB,Balitsky,Kaidalov} and references therein), valid
in a certain high--energy range ($\ldots$ but up to what energy?).
Immediately the following question arises: why our approach, which was
formulated so to give the really asymptotic large--$s$ behaviour of
scattering amplitudes and total cross sections, is also able to reproduce
pre--asymptotic behaviours [violating the FLM bound (\ref{FLM})] like the
one in (\ref{pomeron})?
The answer is clearly that the {\it extra assumptions}, i.e., the {\it models},
which one implicitly or explicitly assumes in the calculation of the
Euclidean correlation functions $\widetilde{\cal C}_E$, play a fundamental
role in this respect. For example, in our approach, developed in the previous
section, we have {\it assumed} that the singularities of the
even--signatured partial--wave amplitudes $A_l^+(t)$ in the complex
$l$--plane (at a given $t$) are only simple poles in $l = \sigma_n^+(t)$.
Every model has its own {\it limitations}, which reflect in the variety of
answers in the literature: someone finds constant cross sections, some other
finds a {\it soft--pomeron} behaviour, some other finds a {\it hard--pomeron}
behaviour $\ldots$
(And maybe the true asymptotic behaviour is $\log^2(s/s_0)$!?).
Unfortunately these {\it limitations} are often out of control, in the sense
that no one knows exactly what is losing due to these approximations.
This is surely a crucial point which, in our opinion, should be further
investigated in the future.

A great help could be provided by a direct lattice calculation
of the loop--loop Euclidean correlation functions \cite{lattice}, whose
analytic continuation to the Minkowskian correlators should furnish (in the
high--energy limit $\chi \to +\infty$) the {\it true} asymptotic behaviour.
Clearly a lattice approach can at most give (after having overcome a lot of
technical difficulties) only a discrete set of $\theta$--values for the
above--mentioned functions, from which it is clearly impossible (without some
extra assumption on the interpolating continuous functions) to get, by the
analytic continuation $\theta \to -i \chi$, the corresponding Minkowskian
correlation functions (and, from this, the elastic scattering amplitudes and
the total cross sections).
However, the lattice approach could provide a criterion to
investigate the goodness of a given existing analytic model (such as:
Instantons, SVM, AdS/CFT, BFKL and so on $\ldots$) or even to open the way
to some new model, simply by trying to fit the lattice data with the
considered model.\\
This would surely result in a considerable progress along this line
of research.

\section*{Acknowledgments}

The author is extremely grateful to O. Nachtmann for useful and enlightening
discussions.

\newpage



\newpage

\noindent
\begin{center}
{\large\bf FIGURE CAPTIONS}
\end{center}
\vskip 0.5 cm
\begin{itemize}
\item [\bf Fig.~1.] The space--time configuration of the two Wilson
loops ${\cal W}_1$ and ${\cal W}_2$ entering in the expression for
the dipole--dipole elastic scattering amplitude in the high--energy limit.
\item [\bf Fig.~2.] The analyticity domain of the function
$\overline{\cal G}_E$ in the complex variable $\theta$.
\item [\bf Fig.~3.] The analyticity domain of the function
$\overline{\cal G}_M$ in the complex variable $\chi$.
\end{itemize}

\newpage

\pagestyle{empty}

\centerline{\large\bf Figure 1}
\vskip 4truecm
\begin{figure}[htb]
\vskip 4.5truecm
\includegraphics{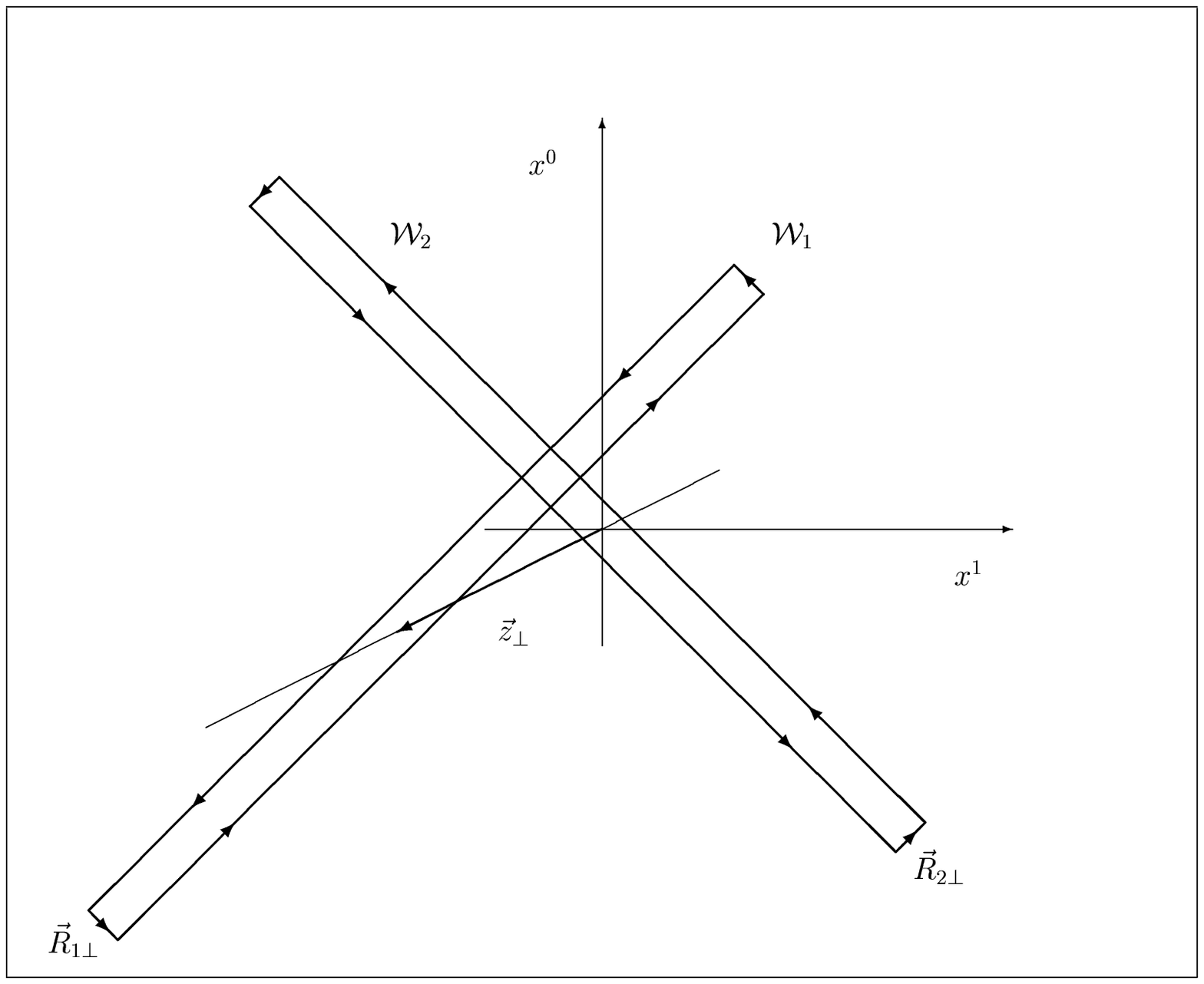}
\end{figure}
\vskip 9.5 cm
\begin{itemize}
\item [\bf Fig.~1.] The space--time configuration of the two Wilson
loops ${\cal W}_1$ and ${\cal W}_2$ entering in the expression for
the dipole--dipole elastic scattering amplitude in the high--energy limit.
\end{itemize}

\newpage

\centerline{\large\bf Figure 2}
\vskip 4truecm
\begin{figure}[ht]
\centering
\includegraphics[height= 0.7\textwidth, width=1\textwidth]{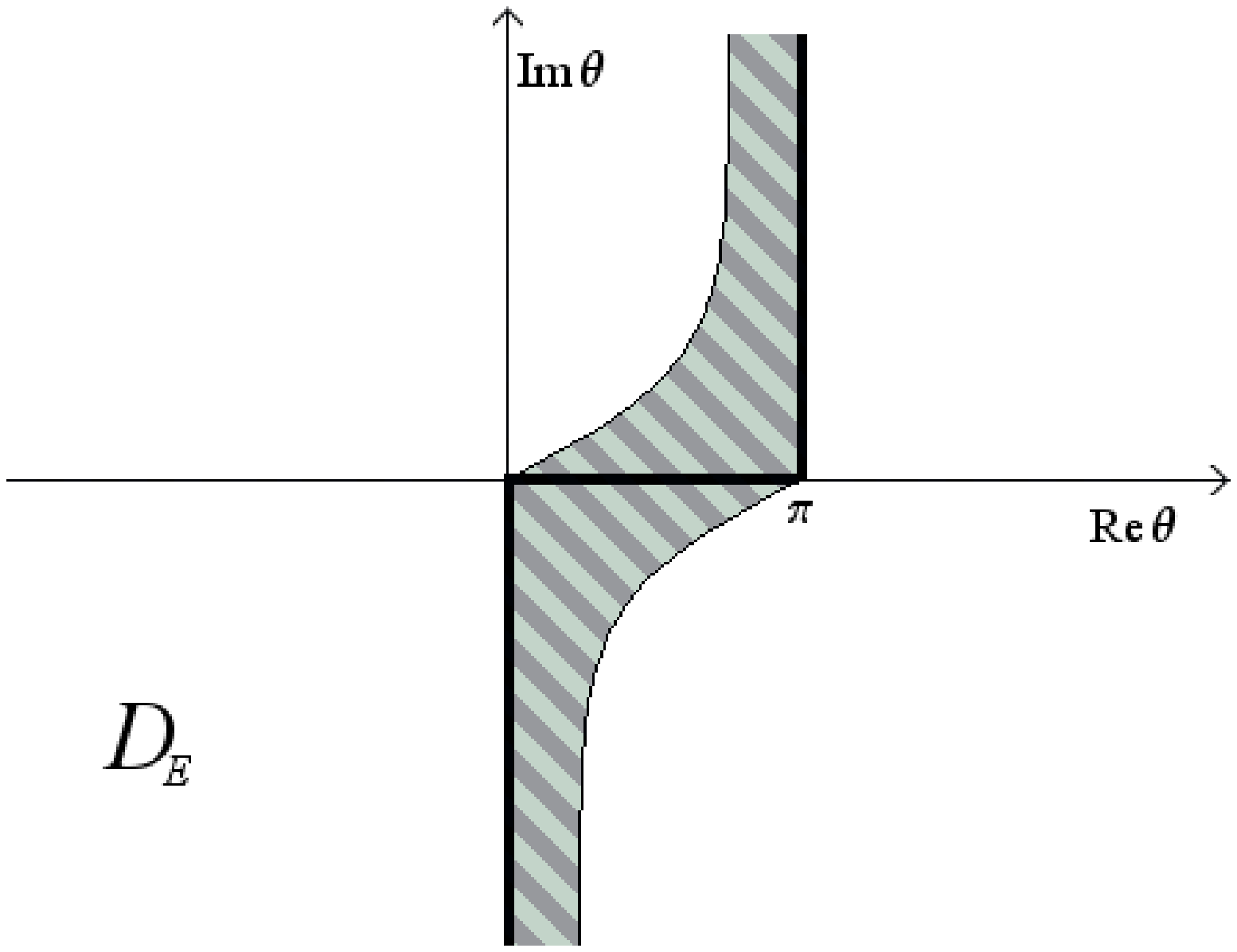}
\label{fig2}
\end{figure}
\begin{itemize}
\item [\bf Fig.~2.] The analyticity domain of the function
$\overline{\cal G}_E$ in the complex variable $\theta$.
\end{itemize}

\newpage

\centerline{\large\bf Figure 3}
\vskip 4truecm
\begin{figure}[ht]
\centering
\includegraphics[height= 0.7\textwidth, width=1\textwidth]{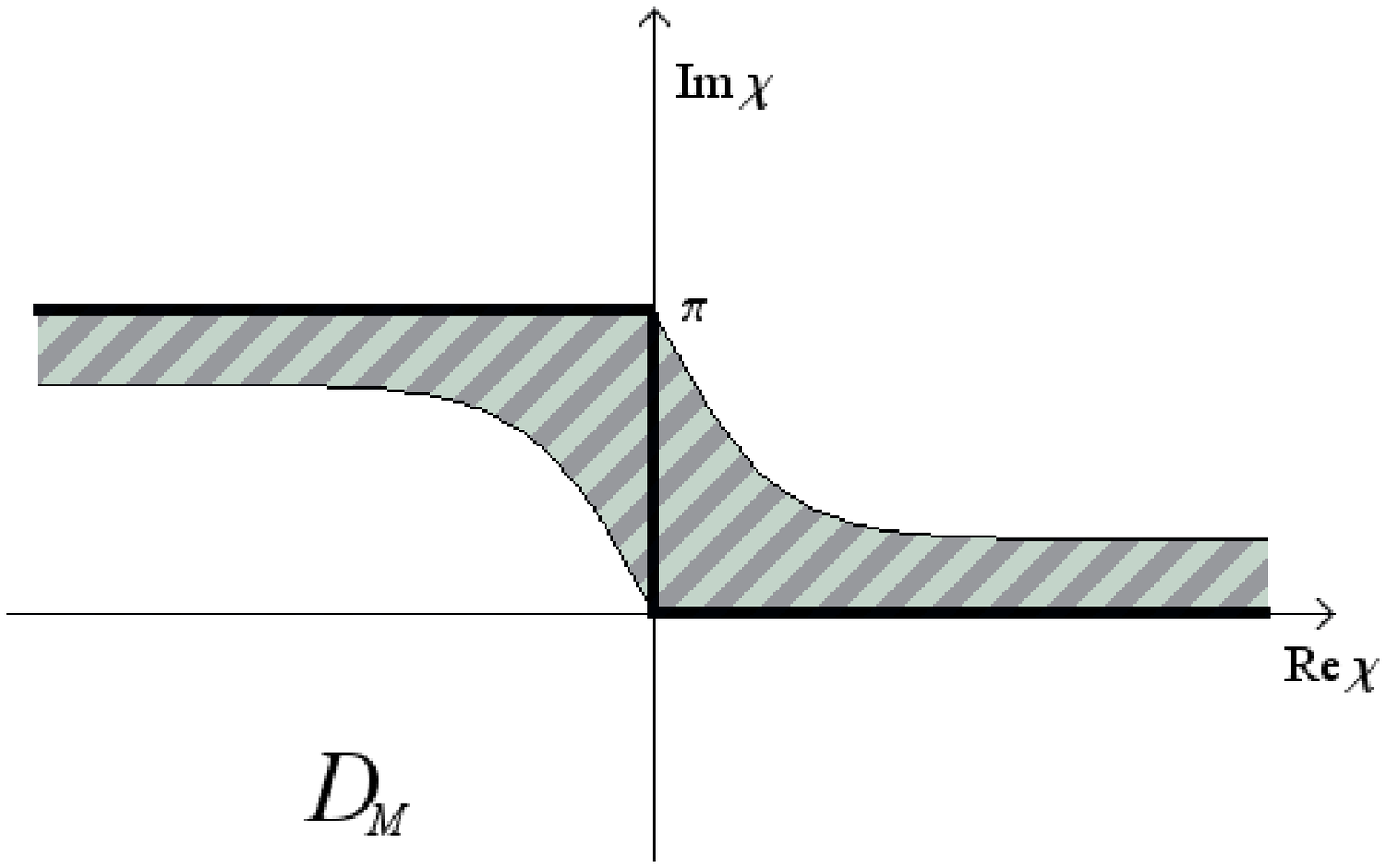}
\label{fig3}
\end{figure}
\begin{itemize}
\item [\bf Fig.~3.] The analyticity domain of the function
$\overline{\cal G}_M$ in the complex variable $\chi$.
\end{itemize}

\end{document}